# Machine Learning for Classifying Vascular Age Using the Second Derivative of the PPG Signal

Corin Otesteanu*
*Biomed. Mobile Health Tech. Lab*
*ETH Zürich*
Zürich, CH
corin.otesteanu@hest.ethz.ch

Gianluca Diana*
*Dept. of Engineering*
*University of Palermo*
Palermo, Italy
gianluca.diana@unipa.it

Francesco Scardulla
*Dept. of Engineering*
*University of Palermo*
Palermo, Italy
francesco.scardulla@unipa.it

Silvia Puleo
*Dept. of Engineering*
*University of Palermo*
Palermo, Italy
silvia.puleo01@unipa.it

Salvatore Pasta
*Dept. of Engineering*
*University of Palermo*
Palermo, Italy
salvatore.pasta@unipa.it

Leonardo D'acquisto
*Dept. of Engineering*
*University of Palermo*
Palermo, Italy
leonardo.dacquisto@unipa.it

Carlo Menon
*Biomed. Mobile Health Tech. Lab*
*ETH Zürich*
Zürich, CH
carlo.menon@hest.ethz.ch

*These authors contributed equally to this work.

***Abstract*— The second derivative of the photoplethysmographic signal features characteristic peaks that provide information on vascular properties. Our analysis was conducted on deflection wave features of the second derivate of PPG signals from human volunteers aged between 26 to 63 years. We show that machine learning models can leverage these features to accurately distinguish subjects based on their age group in a binary classification approach, achieving a balanced accuracy of 80% and an AUC of 0.88 on unseen data, and 100% accuracy on a sample-level split. When including the ratios of deflection wave peaks in our feature set, accuracies of up to 89% and AUC of 0.95 were achieved. Our feature plot illustrates clear clustering between younger and older volunteers, specifically when using the peak ratios, confirming its discriminatory power and its significant potential as a non-invasive index for assessing vascular age. This study demonstrates that second derivative PPG peak ratios, together with the peak amplitudes, are a more robust marker of vascular age than the peak amplitudes alone.**

***Keywords—machine learning, photoplethysmography (PPG), acceleration plethysmogram (APG), wearable devices, optical sensors, signal processing, vascular aging, ai***

## I. Introduction

Assessing the health of the vascular system is a key objective in the prevention and identification of cardiovascular diseases [1]. Among the various indicators, vascular age is a parameter that reflects the degree of aging of the arteries [2], which often differs from the chronological age of the subject [3], and the possibility of estimating this parameter in a non-invasive way is of great scientific importance.

In this context, photoplethysmography (PPG) has proven to be a simple, non-invasive, and low costs optical technique capable of detecting volumetric changes in blood in the microvascular bed. These changes are due to the pressure wave generated by the heart, which causes a transient increase in peripheral blood volume, followed by an outflow during cardiac relaxation within the cardiac cycle [4]. PPG sensors, consisting of a light source and a photodetector, can be easily integrated into wearable devices, such as smartwatches and smartbands, and allow continuous monitoring of physiological parameters such as heart rate and oxygen saturation [5]. In addition to these established applications, the PPG signal offers a still largely unexplored potential for the analysis of vascular function and the estimation of arterial stiffness [6].

One of the most effective and recent methods for extracting morphological information from the PPG signal is the analysis of its second derivative (SDPPG), also known as photoplethysmographic acceleration (APG) [7]. This mathematical transformation highlights rapid changes in the pulse wave, emphasizing five characteristic deflections (a, b, c, d, e) that correspond to specific phases of the cardiac cycle [8]. By combining these parameters, it is possible to extract numerous cardiovascular health indices [9]. In particular, the ratio between the amplitudes of waves d and a (d/a) has been proposed in several studies as an indicator of arterial stiffness and vascular aging [10-11].

To the best of the authors' knowledge, although machine learning has been applied to estimate chronological age from PPG signal morphology [12-13], no previous study has focused on leveraging the second derivative and its characteristic deflection peaks for vascular age assessment. In the present study, we propose an innovative approach based on machine learning techniques to model the relationship between the morphological characteristics of the APG signal and vascular age.

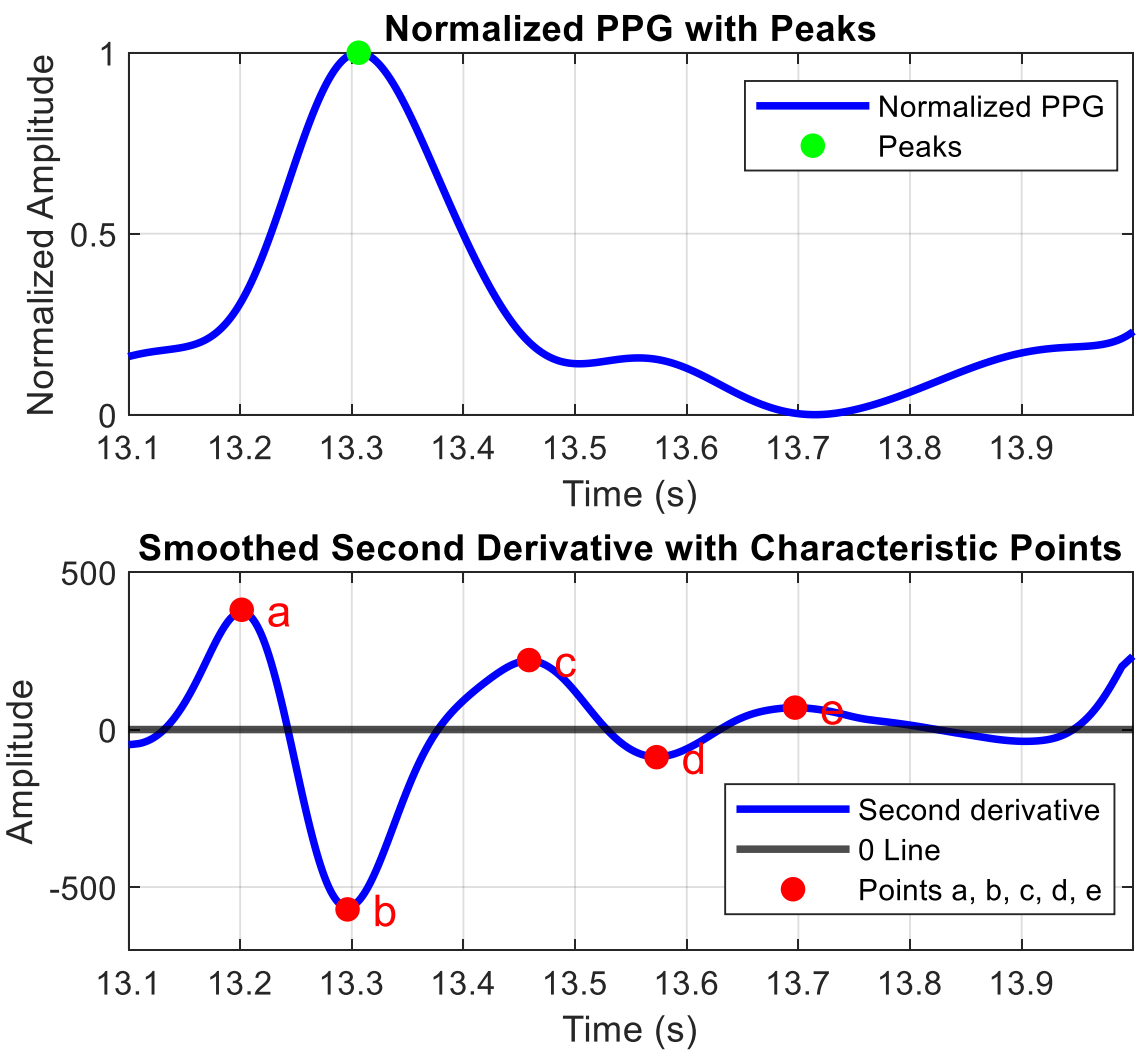

Fig. 1. PPG waveform with corresponding APG waveform and deflection waves.

## II. METHODS

### A. Data properties

The dataset consists in PPG waveforms acquired from the index finger of nine healthy volunteers (6 male, 3 female) aged 26-63 years with no known cardiovascular conditions. PPG signals were obtained using a 520 nm sensor (DFRobot) and digitized at 5 kHz using an NI USB-6009 acquisition system (National Instruments). For each participant, the acceleration photoplethysmogram (APG) was derived by computing the second derivative of the PPG waveform and the five characteristic deflections (a, b, c, d, e waves) were identified and averaged across each acquisition, resulting in a total of five samples per person. A typical PPG waveform with its corresponding APG and deflection waves is shown in Fig. 1.

Participants were assigned to a binary class based on their age group. Class 0 consisted of five individuals, aged up to 40, while class 1 consisted of four individuals above the age of 40. Although vascular aging is continuous, binary classification was adopted as an initial approach, with a threshold of 40 years approximating half the average life expectancy.

### B. Feature representation strategies

We explored three different feature representation strategies derived from the five characteristic deflections of the APG signal. The first approach, referred to as the *raw feature* set, utilized the a, b, c, d, e peaks. The second approach, termed the *ratio feature* set, considered the b/a and d/a ratio. These ratios were chosen because they have been used in previous studies utilizing the PPG-derived acceleration plethysmogram for applications such as blood pressure estimation [14]. Moreover, the age-related increase in arterial stiffness, widely documented in literature. Finally, a *combined feature* set was constructed by integrating both the raw features and the ratio features.

The strategies were quantitatively compared in their ability to discriminate between the two classes using machine learning classifiers.

### C. Training characteristics

Training was performed by splitting the data into training, validation, and test sets. Two training strategies were compared. In the first strategy, the data were split at the sample level, and models were trained using 5-fold cross-validation. In the second strategy, the data were split at the patient level to prevent samples from the same subject appearing in different sets, and models were trained using a leave-one-subject-out (LOSO) 9-fold cross-validation approach.

We evaluated several machine learning models for the classification tasks. Classical algorithms included support vector machines (SVMs), and logistic regression variants, ensemble tree-based methods, such as random forest and XGBoost. K-nearest neighbors (k-NN) was included as a non-parametric approach. In addition, we applied TabPFN, a transformer-based model pre-trained to perform tabular data classification [15].

### D. Evaluation metrics

To evaluate model performance, we computed the area under the receiver operating characteristic curve (AUC), sensitivity, specificity, and balanced accuracy. Balanced accuracy was defined as the mean of the proportion of correct classifications for each class and accounted for the small class imbalance. All evaluation metrics are reported as mean ± standard deviation. In the LOSO evaluation, each test fold contains only a single subject, making individual performance metrics unreliable. Therefore, predictions were pooled across all nine subjects, and a bootstrap procedure was used to compute the standard deviations of performance metrics. In each of the 1000 bootstrap iterations, a sample of the same size as the held-out subject's data was drawn with replacement.

## III. RESULTS AND DISCUSSION

### A. Classification results

At the sample level split, seven out of the eight tested models achieved a balanced accuracy of over 90%, with the transformer based TabPFN model achieving 100% in every evaluation metric (Table I).

TABLE I. COMPARISON OF MACHINE LEARNING MODELS PERFORMANCE USING THE SAMPLE LEVEL CV AND RAW FEATURE SET.

| Model | Balanced Accuracy | AUC | Sensitivity | Specificity |
|---|---|---|---|---|
| Random Forest | 0.96 ± 0.06 | 1.00 ± 0.00 | 0.95 ± 0.10 | 0.96 ± 0.08 |
| SVM | 0.80 ± 0.00 | 0.96 ± 0.02 | 1.00 ± 0.00 | 0.60 ± 0.00 |
| KNN | 0.91 ± 0.05 | 0.97 ± 0.02 | 0.90 ± 0.12 | 0.92 ± 0.10 |
| Logistic Reg | 0.93 ± 0.05 | 0.98 ± 0.04 | 0.95 ± 0.10 | 0.92 ± 0.10 |
| Ridge | 0.96 ± 0.05 | 0.99 ± 0.02 | 1.00 ± 0.00 | 0.92 ± 0.10 |
| Lasso | 0.93 ± 0.05 | 0.98 ± 0.04 | 0.95 ± 0.10 | 0.92 ± 0.10 |
| XGBoost | 0.95 ± 0.06 | 0.97 ± 0.05 | 0.90 ± 0.12 | 1.00 ± 0.00 |
| TabPFN | 1.00 ± 0.00 | 1.00 ± 0.00 | 1.00 ± 0.00 | 1.00 ± 0.00 |

Table II illustrates the performance of our models on the subject level split where results are reported on unseen data. Average model performance dropped by 20% in balanced accuracy, from an average of 93 ± 4% in accuracy in the sample level split to an average of 72.9 ± 6% in the subject level split. Only two models achieved an average accuracy of 80% or above, Lasso (0.86) and TabPFN (0.8). The drop in performance can be partly attributed to the fact that for the sample level split, separate samples acquired from the same participants acquired at different time points were used in the training set. It is possible that the models learned participant-specific characteristics in addition to generalizable age-related patterns, leading to overly optimistic performance estimates. The drop in performance can be attributed also to the limited size of the dataset. Excluding subjects positioned near the decision boundary during training likely impaired the model's ability to generalize. Specifically, in our data one participant was 38 years old and assigned to class 0, while another was 45 years old and assigned to class 1.

TABLE II. COMPARISON OF MACHINE LEARNING MODELS PERFORMANCE ON UNSEEN SUBJECTS AND RAW FEATURE SET.

| Model | Balanced Accuracy | AUC | Sensitivity | Specificity |
|---|---|---|---|---|
| Random Forest | 0.72 ± 0.06 | 0.72 ± 0.08 | 0.77 ± 0.06 | 0.85 ± 0.09 |
| SVM | 0.70 ± 0.06 | 0.70 ± 0.08 | 0.62 ± 0.09 | 0.80 ± 0.10 |
| KNN | 0.68 ± 0.06 | 0.67 ± 0.08 | 0.70 ± 0.07 | 0.75 ± 0.11 |
| Logistic Reg | 0.76 ± 0.07 | 0.74 ± 0.09 | 0.78 ± 0.08 | 0.80 ± 0.10 |
| Ridge | 0.69 ± 0.07 | 0.65 ± 0.09 | 0.78 ± 0.07 | 0.65 ± 0.11 |
| Lasso | 0.86 ± 0.06 | 0.84 ± 0.08 | 0.93 ± 0.04 | 0.80 ± 0.10 |
| XGBoost | 0.62 ± 0.06 | 0.60 ± 0.08 | 0.67 ± 0.08 | 0.65 ± 0.10 |
| TabPFN | 0.80 ± 0.05 | 0.80 ± 0.06 | 0.88 ± 0.05 | 1.00 ± 0.00 |

### B. *Feature representation strategies*

Average feature importance was computed across all models using the scikit-learn framework. For tree-based methods, importance was derived from the mean decrease in impurity, while for linear models it was estimated from the absolute magnitude of the learned coefficients. As shown in Fig. 2a, the most informative features from the raw feature set corresponded to the d, b, and a wave deflection peaks, with an importance above 10%. Fig. 2b illustrates that in addition to the d and b waves, the d/a and b/a ratio also ranked above 10% in importance.

TABLE III. COMPARISON OF MACHINE LEARNING MODEL PERFORMANCE ON UNSEEN SUBJECTS, USING THE RATIO FEATURE SET.

| Model | Balanced Accuracy | AUC | Sensitivity | Specificity |
|---|---|---|---|---|
| Random Forest | 0.80 ± 0.07 | 0.78 ± 0.09 | 0.77 ± 0.08 | 0.80 ± 0.10 |
| SVM | 0.78 ± 0.06 | 0.71 ± 0.10 | 0.76 ± 0.09 | 0.55 ± 0.11 |
| KNN | 0.71 ± 0.07 | 0.67 ± 0.10 | 0.79 ± 0.08 | 0.65 ± 0.12 |
| Logistic Reg | 0.51 ± 0.03 | 0.09 ± 0.07 | 0.42 ± 0.10 | 0.05 ± 0.04 |
| Ridge | 0.78 ± 0.06 | 0.71 ± 0.10 | 0.67 ± 0.11 | 0.55 ± 0.11 |
| Lasso | 0.78 ± 0.06 | 0.71 ± 0.10 | 0.77 ± 0.09 | 0.55 ± 0.12 |
| XGBoost | 0.80 ± 0.07 | 0.78 ± 0.09 | 0.79 ± 0.09 | 0.80 ± 0.10 |
| TabPFN | 0.80 ± 0.07 | 0.78 ± 0.09 | 0.85 ± 0.07 | 0.80 ± 0.10 |

TABLE IV. COMPARISON OF MACHINE LEARNING MODEL PERFORMANCE ON UNSEEN SUBJECTS, USING THE COMBINED FEATURE SET.

| Model | Balanced Accuracy | AUC | Sensitivity | Specificity |
|---|---|---|---|---|
| Random Forest | 0.72 ± 0.06 | 0.71 ± 0.08 | 0.86 ± 0.06 | 0.80 ± 0.10 |
| SVM | 0.70 ± 0.06 | 0.70 ± 0.08 | 0.62 ± 0.09 | 0.80 ± 0.10 |
| KNN | 0.68 ± 0.06 | 0.67 ± 0.08 | 0.70 ± 0.07 | 0.75 ± 0.11 |
| Logistic Reg | 0.76 ± 0.07 | 0.74 ± 0.09 | 0.84 ± 0.06 | 0.80 ± 0.10 |
| Ridge | 0.71 ± 0.07 | 0.68 ± 0.09 | 0.80 ± 0.07 | 0.70 ± 0.11 |
| Lasso | 0.86 ± 0.06 | 0.84 ± 0.08 | 0.93 ± 0.04 | 0.80 ± 0.10 |
| XGBoost | 0.65 ± 0.07 | 0.64 ± 0.09 | 0.71 ± 0.08 | 0.70 ± 0.11 |
| TabPFN | 0.89 ± 0.06 | 0.87 ± 0.07 | 0.95 ± 0.03 | 0.85 ± 0.09 |

Class separability for each feature strategy was evaluated visually by projecting the features onto the first two principal components using principal component analysis (PCA) and plotting the results as a 2D scatter plot, Fig. 2. For this visualization, we used age group labels rather than the binary labels used during model training. This choice was made to highlight cluster separability based on age. The PCA of the raw feature set showed partial separation of age groups, although inter-cluster mixing was present (Fig. 3a). In contrast, the ratio feature set exhibited clearer separation (Fig. 3b), including a well-defined transitional region reflecting the shift from younger to older participants. This aligns with our hypothesis that individuals near the age threshold are more difficult to classify. The transitional region in the PCA plots shows that these edge cases, when absent from the training set, are more prone to misclassification. When both sets were combined, a separation of classes was still visible (Fig. 3c), however the transitional region was less clear. This is likely due to the predominance of the raw features in the variance structure; since PCA seeks to maximize variance, the higher-dimensional raw features outweigh the ratio features.

Evaluation of the models on the ratio feature set and the combined feature set shows an improvement over the raw features, consistent with the patterns observed in the PCA plots. The ratio feature set achieved an average model accuracy of 0.745 ± 0.1, while the combined feature set reached 0.746 ± 0.086. Although the average accuracies are similar, the combined feature set had a positive to neutral effect on all models, whereas the ratio feature set enhanced most models but led to a decrease in performance for two of them. Notably, the ratio feature approach had a negative to neutral effect on the previous two best performing models, the Lasso achieving 0.78 (-0.06%) and the TabPFN 0.80 (+0%). The combined feature approach exerted a neutral to positive effect on these models, which ranked again as the top two. In this evaluation, Lasso achieved an accuracy of 0.86 (+0%) and TabPFN an accuracy of 0.89 (+9%).

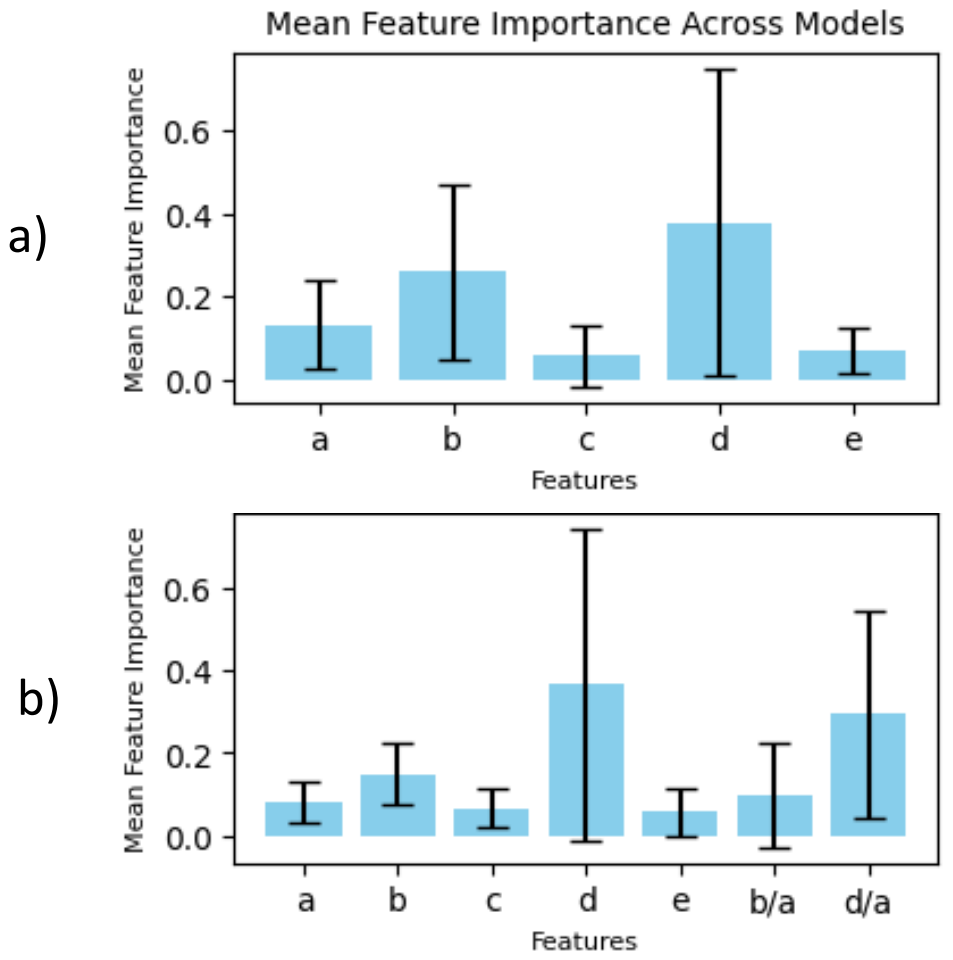

Fig. 2. Mean feature importance across models. Feature importance for the raw feature (a) and combined feature set (b). Error bars represent the standard deviation across models.

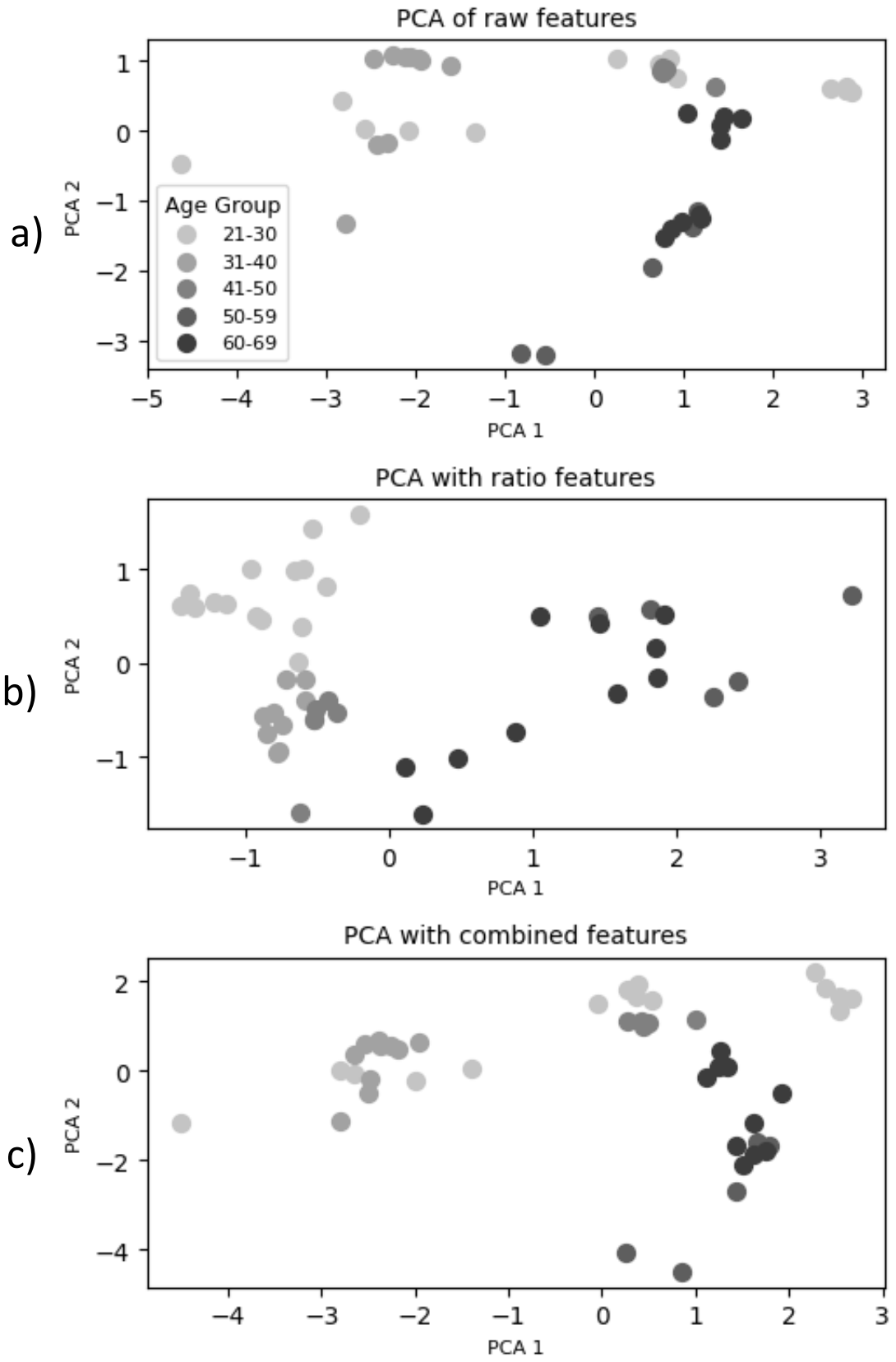

Fig. 3. PPG features from the raw set (a–e), the ratio set, and the combined set were projected onto the first two principal components. Each point represents a sample, and colors indicate patient age group.

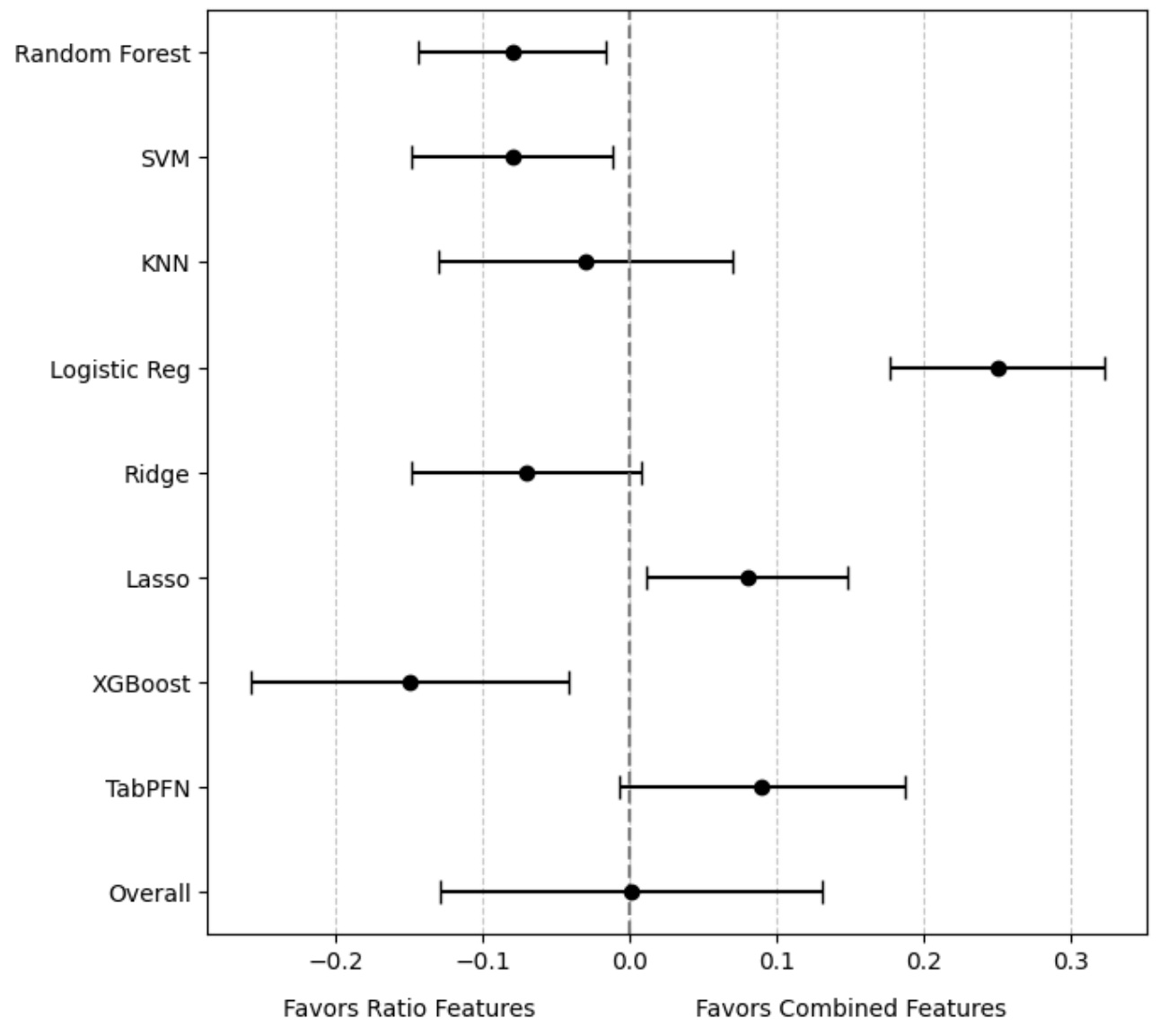

Fig. 4. Forest Plot of Accuracy Improvements Between Combined and Ratio Feature Approaches Across Eight Machine Learning Models

Given the limited dataset of nine subjects, the misclassifications of the two best performing models involved one to two participants, depending on the model, specifically those participants close to the age threshold. These results indicate that morphological features derived from the second derivative of the PPG signal can serve as reliable markers of vascular aging. The importance of the d/a ratio in relation to arterial stiffness was previously reported in literature, likely contributing to the improved performance of our model when included in the feature set [11].

Fig. 4 presents a forest plot summarizing the mean accuracy difference between the combined features and ratio features approaches for each model, with the overall mean displayed at the bottom. The x-axis represents the mean improvement in accuracy, with negative values indicating a preference for the ratio features approach and positive values favoring the combined features approach. Error bars reflect the standard deviation of improvements.

The overall mean difference between the approaches was $0.13 \pm 0.13\%$, with a one-sided t-test indicating no statistically significant difference between approaches. Nevertheless, because the combined feature approach produced neutral to positive effects across all models, it is the preferred approach.

## IV. Conclusion

This study demonstrates that morphological features derived from the second derivative of the PPG signal, particularly deflection wave ratios, can serve as reliable markers of vascular aging. Using machine learning approaches, we showed that both raw APG peaks and their ratios contain discriminatory information, but that the inclusion of ratio features significantly enhances model robustness and separability between age groups. Notably, the combined feature set consistently improved performance across models, with achieving accuracies of 0.86 and 0.89, respectively, on unseen data. These findings highlight the potential of PPG-based vascular biomarkers to complement existing cardiovascular risk assessments and to be integrated into wearable technologies for continuous vascular health monitoring. Given the small sample size (n = 9 healthy subjects), the results should be viewed as proof-of-concept. Future work on larger and more diverse cohorts will be essential to validate the generalizability of APG features for predicting vascular age in real-world settings.

## References

[1] Vatner SF, Zhang J, Vyzas C, Mishra K, Graham RM and Vatner DE (2021) Vascular Stiffness in Aging and Disease. *Front. Physiol.* 12:762437. doi: 10.3389/fphys.2021.762437

[2] KA Groenewegen, HM den Ruijter, G Pasterkamp, JF Polak, ML Bots, Sanne AE Peters, Vascular age to determine cardiovascular disease risk: A systematic review of its concepts, definitions, and clinical applications, *European Journal of Preventive Cardiology*, Volume 23, Issue 3, 1 February 2016, Pages 264–274, https://doi.org/10.1177/2047487314566999

[3] Hamczyk, M, Nevado, R, Barettino, A. et al. Biological Versus Chronological Aging: JACC Focus Seminar. JACC. 2020 Mar, 75 (8) 919–930. https://doi.org/10.1016/j.jacc.2019.11.062

[4] Willemet, M., Alastruey, J. Arterial Pressure and Flow Wave Analysis Using Time-Domain 1-D Hemodynamics. *Ann Biomed Eng* **43**, 190–206 (2015). https://doi.org/10.1007/s10439-014-1087-4

[5] Scardulla, F.; Cosoli, G.; Gnoffo, C.; Antognoli, L.; Bongiorno, F.; Diana, G.; Scalise, L.; D'Acquisto, L.; Arnesano, M. Experimental Analysis on the Effect of Contact Pressure and Activity Level as Influencing Factors in PPG Sensor Performance. *Sensors* 2025, *25*, 4477. doi:10.3390/s25144477

[6] Diana, G.; Scardulla, F.; Puleo, S.; Pasta, S.; D'Acquisto, L. Non-Invasive Estimation of Arterial Stiffness Using Photoplethysmography Sensors: An In Vitro Approach. *Sensors* 2025, *25*, 3301. https://doi.org/10.3390/s25113301

[7] Suboh MZ, Jaafar R, Nayan NA, Harun NH and Mohamad MSF (2022) Analysis on Four Derivative Waveforms of Photoplethysmogram (PPG) for Fiducial Point Detection. *Front. Public Health* 10:920946. doi: 10.3389/fpubh.2022.920946

[8] Elgendi M. On the analysis of fingertip photoplethysmogram signals. *Curr Cardiol Rev.* 2012;8(1):14-25. doi:10.2174/157340312801215782

[9] Elgendi M, Jost E, Alian A, Fletcher RR, Bomberg H, Eichenberger U, Menon C. Photoplethysmography Features Correlated with Blood Pressure Changes. *Diagnostics*. 2024; 14(20):2309. https://doi.org/10.3390/diagnostics14202309

[10] Inoue N, Kawakami H, Yamamoto H, et al. Second derivative of the finger photoplethysmogram and cardiovascular mortality in middle-aged and elderly Japanese women. *Hypertens Res*. 2017;40(2):207-211. doi:10.1038/hr.2016.123

[11] Takazawa K, Tanaka N, Fujita M, et al. Assessment of vasoactive agents and vascular aging by the second derivative of photoplethysmogram waveform. *Hypertension*. 1998;32(2):365-370. doi:10.1161/01.HYP.32.2.36

[12] Lombardi, S., Tavernise, F., Francia, P., Bocchi, L. (2024). Analysis of Age-Related Variations in Photoplethysmography: A Machine Learning Approach. In: *Jarm, T., Šmerc, R., Mahnič-Kalamiza*, S. (eds) 9th European Medical and Biological Engineering Conference. EMBEC 2024. IFMBE Proceedings, vol 112. Springer, Cham. https://doi.org/10.1007/978-3-031-61625-9_11

[13] Abrisham, K.P., Alipour, K., Tarvirdizadeh, B., Ghamari, M.: Neural network models for predicting vascular age from PPG signals: A comparative study. *IET Wirel. Sens. Syst*. e12103 (2025). https://doi.org/10.1049/wss2.12103

[14] Hollmann, N., Müller, S., Eggensperger, K., & Hutter, F. (2022). Tabpfn: A transformer that solves small tabular classification problems in a second. *https://doi.org/10.48550/arXiv.2207.01848*

[15] El-Hajj, C., & Kyriacou, P. A. (2021). Cuffless blood pressure estimation from PPG signals and its derivatives using deep learning models. *Biomedical Signal Processing and Control*, *70*, 102984. *https://doi.org/10.1016/j.bspc.2021.102984*